\documentclass[prd,twocolumn,showpacs,floatfix,amsmath,amssymb,floatfix,nofootinbib]{revtex4}
\usepackage{graphicx,dcolumn,booktabs,bm}
\usepackage{longtable,lscape,multirow}
\usepackage{txfonts}
\usepackage{overpic}
\usepackage{amssymb}
\usepackage{indentfirst}
\usepackage{feynmf}   
\usepackage{slashed}  
\usepackage{cases}
\usepackage{color,ulem}
\usepackage[dvipdfm,  
            colorlinks, 
            pdfborder=001,   
            citecolor=blue
            ]{hyperref}

\graphicspath{{Figures/}} 

\begin{document}


\title{Probing charmonium-like state $X(3915)$ through meson photoproduction}
\author{Qing-Yong Lin$^{1,2,3}$}\email{qylin@impcas.ac.cn}
\author{Xiang Liu$^{1,4}$\footnote{corresponding author}}\email{xiangliu@lzu.edu.cn}
\author{Hu-Shan Xu$^{1,2}$}

\affiliation{
$^1$Research Center for Hadron and CSR Physics, Lanzhou University and Institute of Modern Physics of CAS, Lanzhou 730000, China\\
$^2$Institute of Modern Physics, Chinese Academy of Sciences, Lanzhou 730000, China\\
$^3$University of Chinese Academy of Sciences, Beijing, 100049, China\\
$^4$School of Physical Science and Technology, Lanzhou University, Lanzhou 730000, China}

\date{\today}

\begin{abstract}

Inspired by the observation of charmonium-like state $X(3915)$, we explore the discovery potential of the $X(3915)$ production via meson photoproduction process. By investigating the $\gamma p \to J/\psi\omega p$ process including the $X(3915)$ signal and background contributions, we obtain the corresponding information of the cross section, the Dalitz plot and the $J/\psi\omega$ invariant mass spectrum, which is helpful to further experimental study of $X(3915)$ via meson photoproduction.

\end{abstract}

\maketitle

\section{Introduction}\label{sec:intro}

$X(3915)$ is a typical charmonium-like state, which was firstly observed in the $J/\psi\omega$ invariant mass spectrum of $\gamma\gamma\to J/\psi\omega$. It's mass and width are
$M_{X(3915)}=(3915\pm3(\mathrm{stat})\pm 2(\mathrm{syst}))$ MeV and
$\Gamma_{X(3915)}=(17\pm10(\mathrm{stat})\pm3(\mathrm{syst}))$ MeV, respectively.
Since $X(3915)$ comes from the $\gamma\gamma$ fusion process, either $J^P=0^{+}$ or $J^{P}=2^{+}$ is the possible spin-parity assignment to $X(3915)$. Furthermore, Belle's measurement gives $\Gamma_{X(3915)\to \gamma\gamma}\cdot BR(X(3915)\to
J/\psi\omega) = (61\pm 17(\mathrm{stat})\pm 8(\mathrm{syst}))$ eV and $(18\pm
5(\mathrm{stat})\pm 2(\mathrm{syst}))$ eV corresponding to $J^P=0^{+}$ and $J^{P}=2^{+}$, respectively \cite{:2009tx}.

What is its inner structure is a crucial question to reveal the property of $X(3915)$. In Ref. \cite{Liu:2009fe},
$X(3915)$ as the first radial excitation of $\chi_{c0}(1P)$ was proposed by the analysis of mass spectrum and the study of its strong decay behavior. This conventional charmonium explanation to $X(3915)$ was confirmed by the BaBar Collaboration, where BaBar carried out a spin-parity analysis of $X(3915)$, which supports the $J^P=0^+$ assignment due to the $\chi_{c0}(2P)$ resonance \cite{Lees:2012xs}.
However, $X(3915)$ as a $\chi_{c0}(2P)$ state brings us a new problem. A $\chi_{c0}(2P)$ state dominantly decays into $D\bar{D}$ \cite{Liu:2009fe}. Until now the $X(3915)$ signal has been missing in the $D\bar{D}$ invariant mass spectrum given by Belle \cite{Uehara:2005qd} and BaBar \cite{Aubert:2010ab} in $\gamma\gamma\to D\bar{D}$, where the charmonium-like state $Z(3930)$ was reported \cite{Uehara:2005qd,Aubert:2010ab}, which is a $\chi_{c2}(2P)$ state. For solving this problem, the authors in Ref. \cite{Chen:2012wy}
proposed that the $Z(3930)$ enhancement structure may contain two P-wave higher charmonia $\chi_{c0}(2P)$ and $\chi_{c2}(2P)$. The numerical result
shows that this assumption is supported by the analysis of the $D\bar{D}$ invariant mass spectrum and the $\cos\theta$ distribution of $\gamma\gamma\to D\bar{D}$ \cite{Chen:2012wy}. Thus, there does not exist any difficulty to this $\chi_{c0}(2P)$ explanation to $X(3915)$.

Just introduced in the above, $X(3915)$ was only reported in the $\gamma\gamma$ fusion process. It is also natural to ask whether $X(3915)$ can be found in other processes. At present, according to the production mechanisms, all observed charmonium-like states reported in the past decade can be roughly categorized into four groups, which correspond to the $\gamma\gamma$ fusion process, the $B$ meson decay with the kaon emission, the $e^+e^-$ annihilation, the hidden-charm dipion decays of higher charmonia and charmonium-like states.
In the past yeas, there are some theoretical explorations of the production of charmonium-like state by other processes different from the above four mechanisms mentioned above. {For example, Liu, Zhao and Close suggesed that
$Z(4430)^\pm$ can be produced via the meson photoproduction process $\gamma p\to Z(4430)^+ n\to \psi^\prime \pi^+ n$ \cite{Liu:2008qx}. While the authors in Ref. \cite{Ke:2008kf} proposed to study $Z(4430)^\pm$ by the nucleon-antinucleon scattering at the forthcoming PANDA experiment. In Ref. \cite{He:2009yda}, He and Liu discussed the discovery potential of charmonium-like state $Y(3940)$ by the meson photoproduction process $\gamma p\to Y(3940)p$. Very recently, the production of the newly observed charged $Z_c(3900)$ by the photoproduction process $\gamma p\to Z_c(3900)n$ was discussed \cite{Lin:2013mka}. }
It is obvious that  probing the production of $X(3915)$ by different processes from the $\gamma\gamma$ fusion is an interesting research topic.

We notice the $X(3915)\to J/\psi\omega$ decay mode \cite{:2009tx}. Since both $J/\psi$ and $\omega$ in the final state of the $X(3915)\to J/\psi\omega$ decay are vector meson, we suppose that we can carry out the study of the $X(3915)$ production via meson photoproduction. For further testing this idea, in this work we need to choose suitable meson photoproduction process to study the production of $X(3915)$ and discuss the discovery potential of $X(3915)$ through meson photoproduction. In the next section, we illustrate the detailed consideration.

This paper is organized as follows. In Sec. \ref{sec:intro}, we introduce the research status of $X(3915)$ and our motivation of studying the production of $X(3915)$ by meson photoproduction. After introduction, we present the details of whole calculation. And the numerical result will be given in Sections \ref{sec2} and \ref{sec3}. The last section is a short summary.

\section{The production of X(3915) via meson photoproduction}\label{sec2}

\begin{figure}[htb]
\begin{center}
\scalebox{0.9}{\includegraphics[width=\columnwidth]{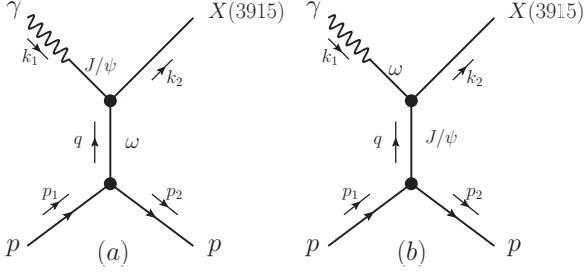}}
\caption{The $\gamma p \to X(3915)p$ process with the $\omega$ (a) and $J/\psi$ (b) meson exchanges.
\label{fig:2to2}}
\end{center}
\end{figure}

Under the assumption of vector meson dominance (VMD) \cite{Bauer:1975bv,Bauer:1975bw,Bauer:1977iq},
$\omega$ or $J/\psi$ can interact with a photon. Thus, by exchanging a vector meson like $\omega$ or $J/\psi$, $X(3915)$ can be produced via the meson photoproduction process $\gamma p \to X(3915)p$, which is shown in Fig. \ref{fig:2to2}.
As discussed in Refs. \cite{Barnes:2006ck,Lin:2012ru} and references therein, the coupling of $\omega$ with the nucleons is obviously stronger than that of $J/\psi$ interacting with the nucleons.
In addition, the mass of the exchanged $J/\psi$ is far larger than that of the exchanged $\omega$ meson.
Thus, the contribution from Fig. \ref{fig:2to2} (a) must be far larger than that from Fig. \ref{fig:2to2} (b), which makes us only consider the contribution from Fig. \ref{fig:2to2} (a) in our study.
Later, we will present the results of the total cross section and differential cross section.

For depicting the interaction of the $\omega$ meson with the nucleons, we adopt the following effective Lagrangian \cite{Downum:2006re}
\begin{eqnarray}
{\cal L}_{NN\omega} &=& -g_{NN\omega} \left(\bar{\Psi}\gamma_\mu \Psi \phi^\mu -
\frac{\kappa_\omega}{4m_N}\bar{\Psi}\sigma_{\mu\nu} \Psi F^{\mu\nu}\right) ,
\end{eqnarray}
where $\Psi$ and $\phi$ denote the fields of the nucleon and the $\omega$ meson, respectively.
We define $F_{\mu\nu}=\partial_{\mu}\phi_{\nu}-\partial_{\nu}\phi_{\mu}$ and $\sigma_{\mu\nu}=i[\gamma_\mu,\gamma_\nu]/2$.
For the $pp\omega$ vertex, there exist two strong coupling constants $g_{NN\omega}$ and $\kappa_\omega$. At present, these coupling constants are not well determined. Different models and approaches gave different theoretical estimates for these two coupling constants \cite{Lin:2012ru}. In Table \ref{tab:omega_couplings}, we list different combinations of the $g_{NN\omega}$ and $\kappa_\omega$ values from different models.
Additionally, a monopole form factor $F_{NN\omega}(q^2)=(\Lambda_\omega^2 - m_\omega^2)/(\Lambda_\omega^2 - q^2)$ is introduced to compensate the off-shell effect of the exchanged $\omega$ meson, where the cutoff $\Lambda_\omega$ is set to be 1.2 GeV \cite{Gasparyan:2003fp}.
\renewcommand{\arraystretch}{1.5}
\begin{table}[htb]
\caption{The theoretical values of the coupling constants $g_{NN\omega}$ and $\kappa_\omega$.
\label{tab:omega_couplings}}
\begin{tabular}{ccc}
\toprule[1pt]
Mechanism/Model & $g_\omega$ & $\kappa_\omega$ \\\midrule[0.5pt]

Paris \cite{Cottingham:1973wt, Lacombe:1980dr} & 12.2 & $-0.12$ \\
Nijmegen \cite{Nagels:1978sc} & 12.5 & $+0.66$ \\
Bonn \cite{Machleidt:2000ge} & 15.9 & 0 \\

Pion photoproduction \cite{Sato:1996gk} & $7-10.5$ & 0 \\

Nucleon EM form factors \cite{Mergell:1995bf} & $20.86\pm0.25$ & $-0.16\pm0.01$ \\

QCD sum rule \cite{Zhu:1999kva} & $18\pm8$ & $0.8\pm0.4$ \\

$^3P_0$ quark model \cite{Downum:2006re} & $-$ & $-3/2$ \\
\multirow{2}{*}{Light meson emission model \cite{Barnes:2010yb}} & $23\pm3$ & 0
\\
& $14.6\pm2.0$ & $-3/2$
\\
\bottomrule[1pt]
\end{tabular}
\end{table}

According to the VMD mechanism \cite{Bauer:1975bv,Bauer:1975bw,Bauer:1977iq}, the Lagrangian describing the interaction between the photon and the vector meson can be expressed as
\begin{equation}
{\cal L}_{V\gamma} = -\frac{em_V^2}{f_V}V_\mu A^\mu,
\end{equation}
where $m_V$ and $f_V$ are the mass and the decay constant of the vector meson $V$, respectively. The decay constant $f_V$ can be determined by the decay $V\to e^{+}e^{-}$, i.e.,
\begin{eqnarray}\label{eqa:vmd}
\frac{e}{f_V} &=& \left[\frac{3\Gamma_{V\to e^+ e^-}m_V^2}{8\alpha |\textbf{k}|^3}\right]^{1/2},
\end{eqnarray}
where $|\textbf{k}|=(m_V^2-4m_e^2)^{1/2}/2\simeq m_V/2$ is the three momentum of electron in the rest
frame of the vector meson. $\alpha = e^2/(4\pi) = 1/137$ is the fine-structure constant.
By $\Gamma_{J/\psi\to e^+e^-}=5.55\pm 0.14\pm 0.02$
keV~\cite{Beringer:1900zz} and Eq. (\ref{eqa:vmd}), one obtains $e/f_{J/\psi}= 0.027$.

Under the assignment of quantum number $J^{PC}=0^{++}$ to $X(3915)$ \cite{Liu:2009fe,Lees:2012xs}, the interaction of $X(3915)$ with $J/\psi\omega$ can be depicted by  \cite{Black:2002ek,Nagahiro:2008mn}
\begin{eqnarray}\label{eqa:ver}
\left\langle J/\psi(k_1)\omega(\epsilon_2,k_2)|X(3915)(\epsilon_1)\right\rangle &=& g_{XJ/\psi\omega}T_{\mu\nu}\epsilon_1^{\mu}\epsilon_2^{\nu}
\end{eqnarray}
with
\begin{eqnarray}
T^{\mu\nu} &=& g^{\mu\nu}k_1\cdot k_2 - k_1^{\nu}k_2^{\mu},\nonumber
\end{eqnarray}
where $\epsilon_1(\epsilon_2)$ is the polarization vector of $J/\psi(\omega)$. The coupling constant $g_{XJ/\psi\omega}$ can be determined by the decay width of $X(3915)\to J/\psi\omega$, i.e.
\begin{eqnarray}\label{gamma-x}
\Gamma\left[X \to \psi \omega\right] &=& \frac{1}{8\pi}\frac{|{\textbf{p}^\prime}|}{m_X^2} |{\cal{M}}[X \to \psi \omega]|^2 \nonumber\\
&=&\left(\frac{g_{X\psi\omega}}{m_X}\right)^2 \frac{ |{\textbf{p}^\prime}|}{16\pi}
 \left[ m_\psi^4 + 2M_\psi^2(2m_\omega^2 - m_X^2) \right. \nonumber \\
 &&\left. + (m_\omega^2-m_X^2)^2\right] ,
\end{eqnarray}
where $\textbf{p}^\prime=[(m_X^2-(m_\psi+m_\omega)^2)(m_X^2-(m_\psi-m_\omega)^2)]^{1/2}/(2m_X)$ denotes the three momentum of the daughter mesons in the parent's center of mass frame. At present, the decay width of $X(3915)\to J/\psi\omega$ has not been well determined in experiment.
In order to discuss the production of $X(3915)$ via photoproduction process, we adopt several typical values of the decay width of $X(3915) \to J/\psi\omega$ as input, i.e., we take $\Gamma(X\to J/\psi\omega)=0.15,\,1.7,\,5.1$ MeV\footnote{The lower and upper limits of theoretical two-photon decay width are 1 keV to 5.47 keV, respectively, which are from different model calculations \cite{Godfrey:1985xj,Munz:1996hb,Hwang:2010iq,Ebert:2003mu,Wang:2007nb,Chen:2013yxa}. Thus, by Belle's measurement $\Gamma_{X(3915)\to \gamma\gamma}\cdot BR(X(3915)\to
J/\psi\omega) = (61\pm 17(\mathrm{stat})\pm 8(\mathrm{syst}))$ eV \cite{:2009tx}, we can obtain the corresponding range of the decay width of $X(3915)\to J/\psi\omega$, which is $0.19\sim1$ MeV. The typical values of $X(3915)\to J/\psi\omega$ taken in this work cover the above theoretical range.}, which correspond to the coupling constants $g_{XJ/\psi\omega}=0.118,\, 0.3956,\,0.6852$ GeV$^{-1}$, respectively.

\begin{figure*}[htbp]
\begin{center}
\begin{tabular}{cc}
\scalebox{0.9}{\includegraphics[width=\columnwidth]{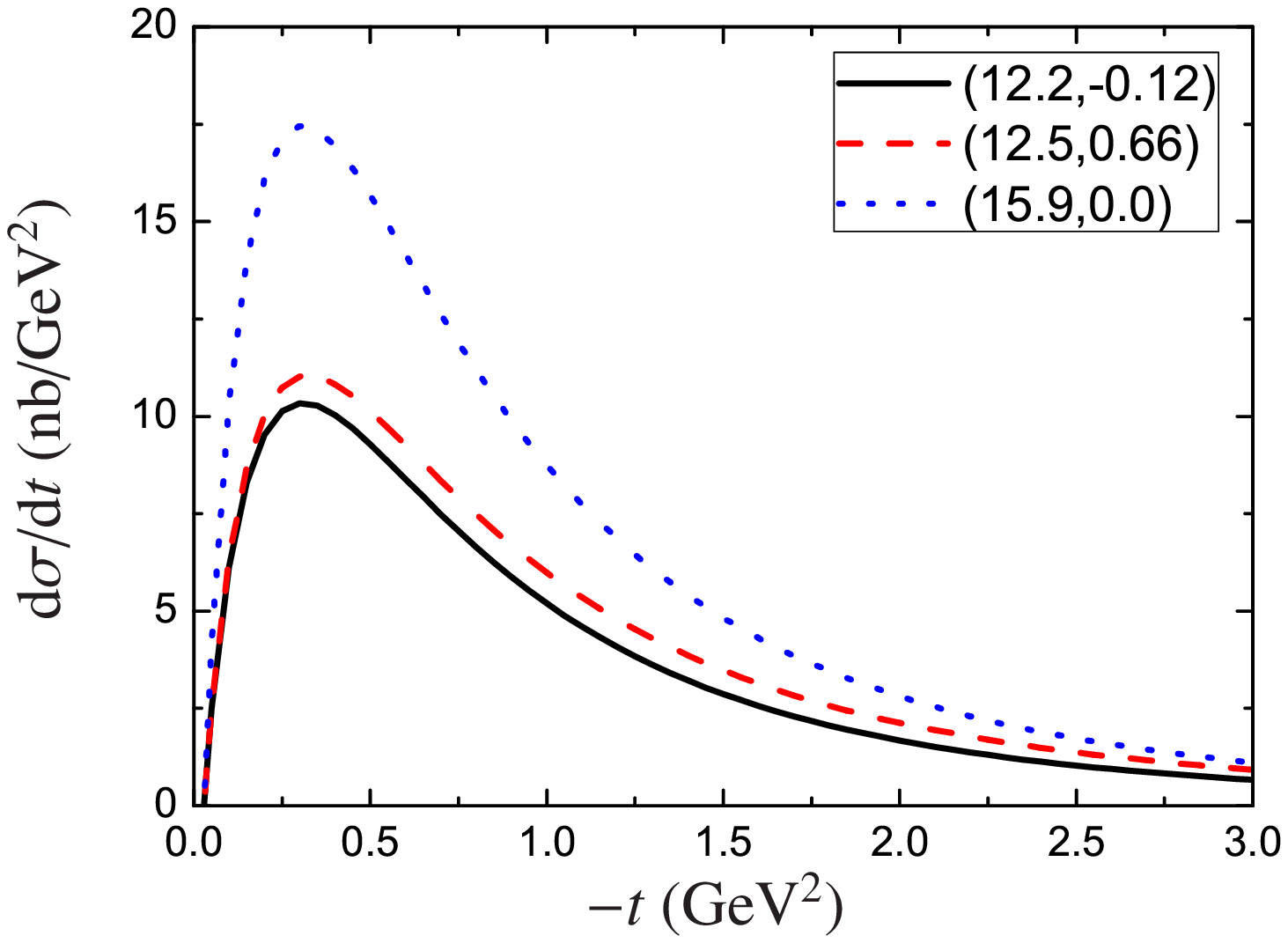}}&\scalebox{0.9}{\includegraphics[width=\columnwidth]{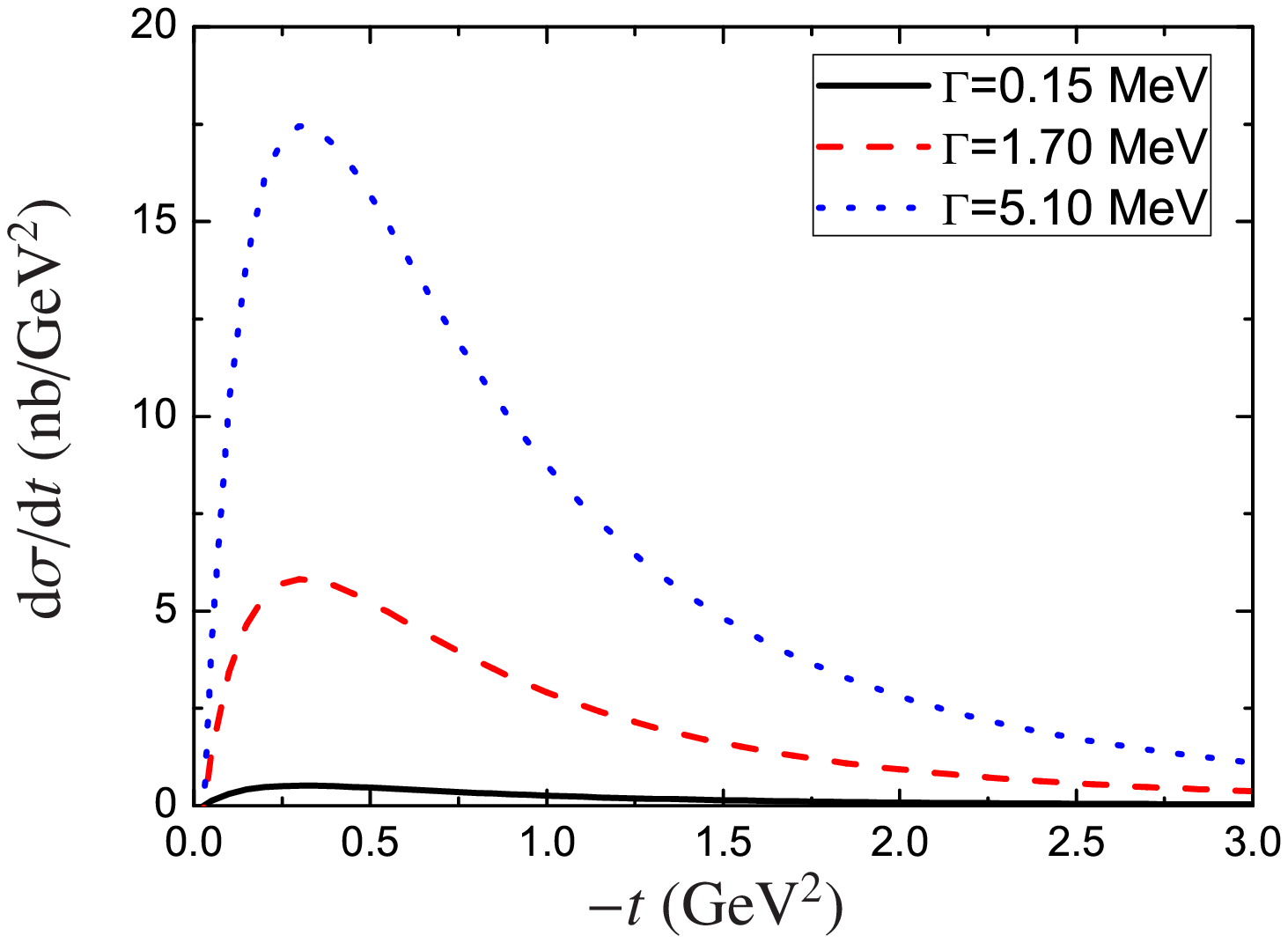}}\\
(a)&(b)\\
\scalebox{0.9}{\includegraphics[width=\columnwidth]{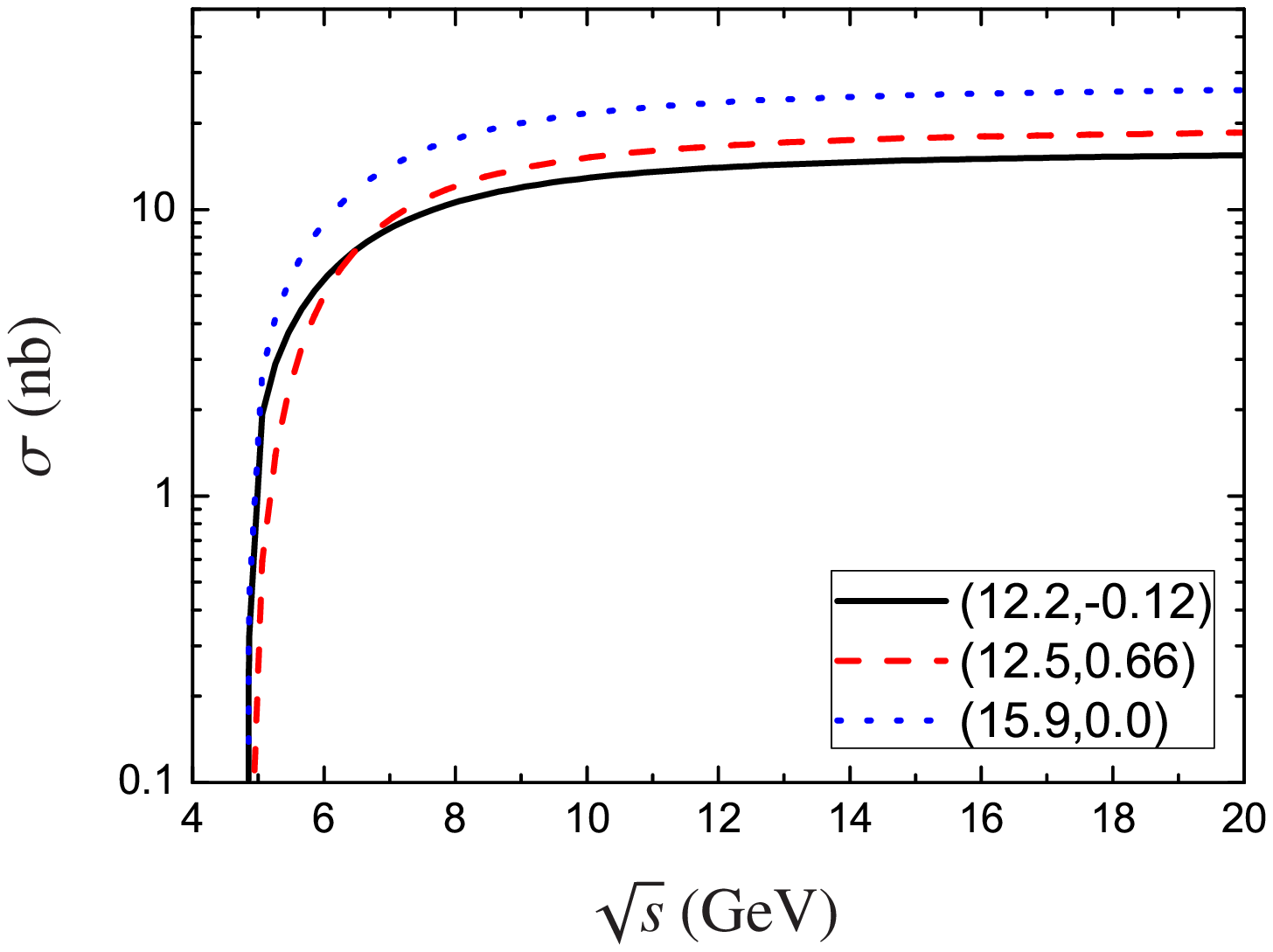}}&\scalebox{0.9}{\includegraphics[width=\columnwidth]{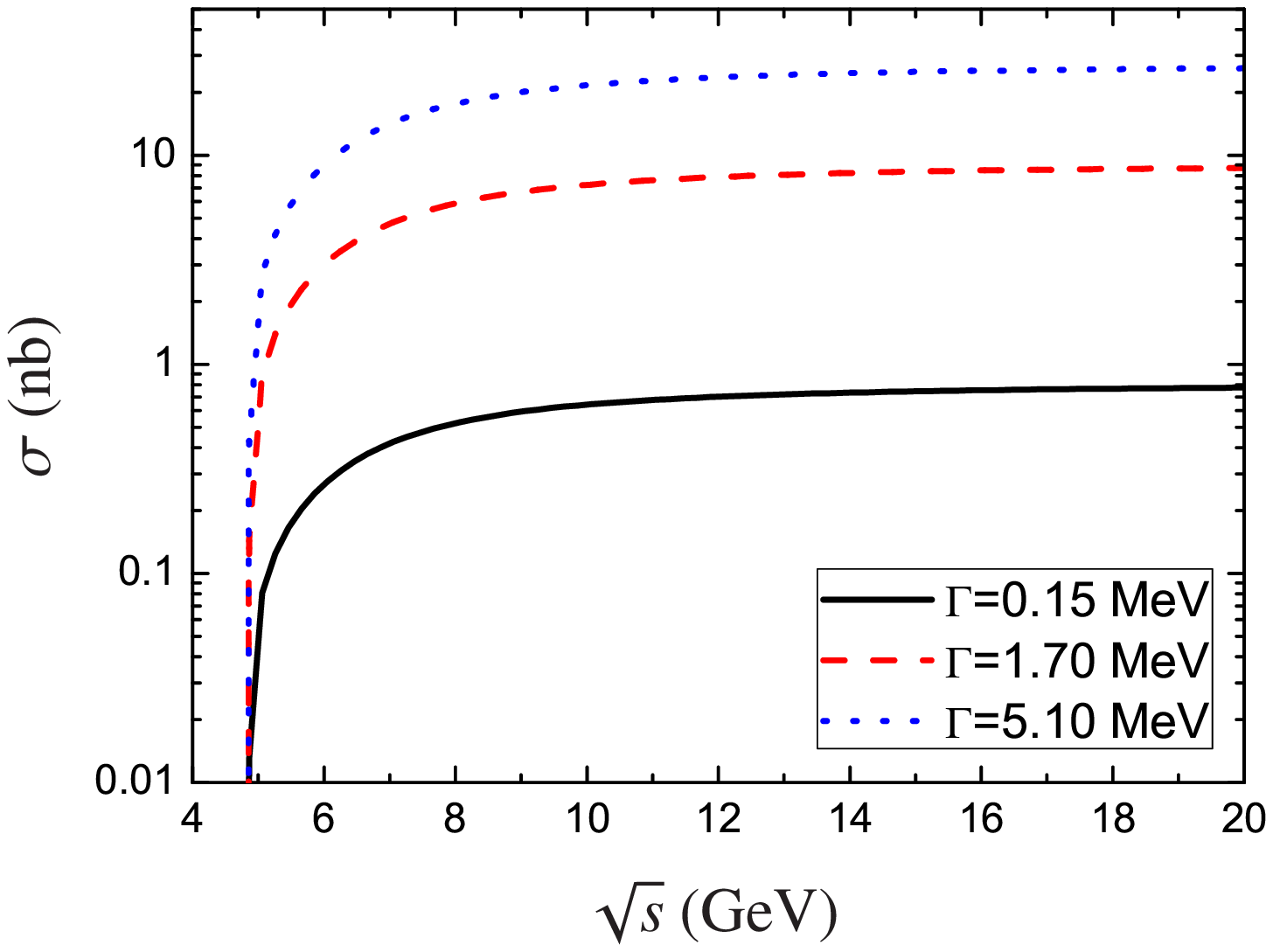}}\\
(c)&(d)\\
\end{tabular}
\end{center}\caption{(Color online). The differential and total cross sections of $\gamma p \to X\mbox{(3915)}p$ dependent on $-t$ and $\sqrt{s}$, respectively. (a) and (c) are the results of the differential and total cross sections with $\Gamma(X(3915)\to J/\psi\omega)=5.1$ MeV and several typical values of the coupling constants $(g_{NN\omega},\kappa_\omega)$, respectively.
(b) and (d) are the results of the differential and total cross sections with the fixed $(g_{NN\omega},\kappa_\omega)=(15.9,0)$ and the typical values of decay width of the $X(3915)\to J/\psi\omega$ decay, respectively.
\label{fig:DSDT2to2}}
\end{figure*}

With the Lagrangians given above, we obtain the transition amplitude for $\gamma p \to X(3915)p$ shown in Fig. \ref{fig:2to2}
\begin{eqnarray}
\mathcal{T}_{fi} &=& \left(\frac{e}{f_\psi}g_{NN\omega}g_{X\psi\omega}\right)
(g^{\mu\nu}k_1\cdot q -k_1^\nu q^\mu)\frac{-g_{\nu\alpha}+q_\nu q_\alpha/m_\omega^2}{q^2-m_\omega^2} \nonumber \\
&& \times  \bar{u}(p_2)\left(\gamma^\alpha + \frac{i\kappa_\omega}{2m_N}\sigma^{\alpha\beta}q_\beta\right)u(p_1) \epsilon_\gamma^\mu \nonumber \\ && \times  F_{NN\omega}(q^2)F_{X\psi\omega}(q^2),
\end{eqnarray}
where we also introduce a monopole form factor $F_{XJ/\psi\omega}(q^2)=(\Lambda_X^2 - m_\omega^2)/(\Lambda_X^2 - q^2)$ to describe the vertex with an off-shell $\omega$ meson and take $\Lambda_X=m_\psi$ \cite{Liu:2008qx}. And then, the unpolarized differential cross section reads as
\begin{equation}\label{eqa:dsdt}
 \frac{d\sigma}{dt} = \frac{1}{64\pi s} \frac{1}{|k_{1cm}|^2}
 \frac{1}{4}\sum\limits_{pol} \left| \mathcal{T}_{fi}  \right|^2
\end{equation}
where $k_{1cm}$ is the photon energy in the center of mass frame of the $\gamma p$ scattering. $s=(k_1 + p_1)^2$ and $t=(k_2-k_1)=q^2$ denote the Mandelstam variables. Just indicated in Eq. \ref{eqa:dsdt}, the total cross section is proportional to $g_{X\psi\omega}^2$.

In Fig. \ref{fig:DSDT2to2}, we show the differential cross section of $\gamma p \to X(3915)p$ dependent on $-t$, where two cases are considered: (1) We present the results corresponding to several $(g_{NN\omega},\kappa_\omega)$ coupling constants. (2) We also list the results with three typical decays widths of $X(3915)\to J/\psi\omega$. While the corresponding dependence of the total cross section of $\gamma p \to X(3915)p$ on $\sqrt{s}$ is given in Fig. \ref{fig:DSDT2to2}. We need to specify that the differential cross section is calculated with the fixed value $\sqrt{s}=9.9$ GeV.

The results of differential cross section show that there is a peak structure at low $-t$ region, which increases rapidly near the threshold and then decreases slowly with the increase of $-t$.
The line shape of the total cross section goes up very rapidly near the threshold, and then increases slowly with the increase of $\sqrt{s}$.

\section{The Background Analysis and Dalitz Plot}\label{sec3}

In the above section, we discussed the production of $X(3915)$ via the $\gamma p \to X(3915)p$ process.
For providing more abundant information of the discovery potential of $X(3915)$ via the meson photoproduction, in the following we perform the background analysis and the Dalitz plot, where we not only consider the $X(3915)$ signal contribution but also include the background contribution.
Since $X(3915)$ can decay into $J/\psi\omega$ \cite{:2009tx}, we analyze the $\gamma p \rightarrow X(3915)p \rightarrow J/\psi\omega p$ reaction in detail.


\begin{figure}[htbp]
\begin{center}
\scalebox{0.9}{\includegraphics[width=\columnwidth]{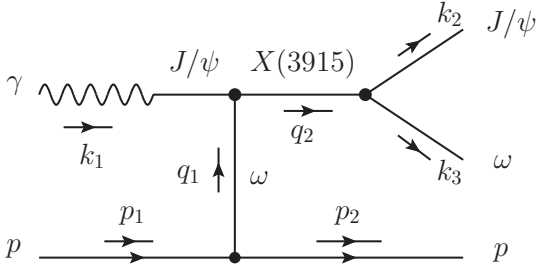}}
\caption{The process $\gamma p \to X(3915)p \to J/\psi\omega p$ via the $\omega$ meson exchange.
\label{fig:2to3s}}
\end{center}
\end{figure}

As the main signal contribution, the $\gamma p \to X(3915)p \to J/\psi\omega p$ reaction with the $\omega$ exchange is described in Fig. \ref{fig:2to3s}.
Adopting the effective Lagrangian approach, we write out the transition amplitude
\begin{eqnarray}\label{trans-sig}
\mathcal{T}_{fi}^{Singal} &=& \left(g_{NN\omega}g_{X\psi\omega}^2 \frac{e}{f_{\psi}}\right)\bar{u}(p_2)(\gamma_\rho+\frac{i\kappa_\omega}{2m_N}\sigma_{\rho\lambda}q_{1\lambda})u(p_1) \nonumber\\
&& \times \left(g_{\nu\alpha}k_2\cdot k_3-k_{2\alpha}k_{3\nu})(g_{\mu\beta}k_1\cdot q_1-k_{1\beta}q_{1\mu}\right) \nonumber\\
&& \times \frac{(-g^{\beta\rho}+q_1^{\beta}q_1^{\rho}/m_\omega^2)}{q_1^2-m_\omega^2}
\frac{\epsilon_{\gamma}^{\mu}\epsilon_{\psi}^{\ast\nu}\epsilon_{\omega}^{\ast\alpha}}{(q_2^2-m_X^2+im_X\Gamma_X)}\nonumber\\
&& \times \left(\frac{\Lambda_\omega^2-m_\omega^2}{\Lambda_\omega^2-q_1^2} \right)\left(\frac{\Lambda_X^2
-m_\omega^2}{\Lambda_X^2-q_1^2} \right)\left(\frac{\Lambda_X^2-m_X^2}{\Lambda_X^2-q_2^2} \right),\label{h0}
\end{eqnarray}
where the cutoff is taken as $\Lambda_X=m_\psi$ \cite{Friman:1995qm}. In Fig. \ref{fig:2to3s}, the relevant
kinematic variables are marked.

\begin{figure}[htbp]
\begin{center}
\scalebox{0.45}{\includegraphics[width=\textwidth]{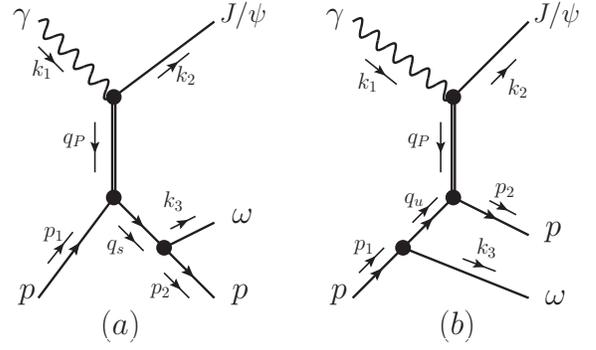}}
\caption{The $\gamma p\to J/\psi\omega p$ process through the Pomeron exchange.
\label{fig:2to3bg}}
\end{center}
\end{figure}

The Pomeron contribution has been widely applied to the study of the diffractive
transitions in Refs. \cite{Donnachie:1987pu,Pichowsky:1996jx,Laget:1994ba}. Since the Pomeron can mediate the long-distant interaction between a confined quark and a nucleon and behaves like an isoscalar photon with $C=+1$, the Pomeron exchange can mainly contribute to the $\gamma p \to J/\psi p \to J/\psi\omega p$ reaction, which provides the background contribution. The concrete description of this process is given in Fig. \ref{fig:2to3bg}.

To describe the Pomeron exchange process shown in Fig. \ref{fig:2to3bg}, we adopt the
formula given in Refs. \cite{Liu:2008qx,Donnachie:1987pu,Pichowsky:1996jx,Zhao:1999af}.
The Pomeron-nucleon coupling can be expressed as
\begin{equation}
F_\mu(\tilde{t})=\frac{3\beta_0(4m_N^2-2.8\tilde{t})}{(4m_N^2-\tilde{t})(1-\tilde{t}/0.7)^2}\gamma_\mu= f(\tilde{t})\gamma_\mu,
\end{equation}
where $\tilde{t}=q_{P}^2=(k_1-k_2)^2$ is the the exchanged Pomeron momentum squared. $\beta_0$ denotes the
coupling constant between a single Pomeron and a light constituent
quark.

As suggested in Ref.~\cite{Zhao:1999af}, an
on-shell approximation with the gauge-fixing scheme
is applied to depict the $\gamma V \mathcal{P}$ vertex, where the equivalent vertex for the
$\gamma\psi\mathcal{P}$ interaction is written as
\begin{eqnarray}
V_{\gamma V\cal{P}} &=& \frac{2\beta_c \times 4\mu_0^2}{(M_{\psi}^2
-\tilde{t})(2\mu_0^2+M_{\psi}^2 -\tilde{t})} T_{\mu\alpha\nu}\epsilon_\gamma^\mu\epsilon_\psi^{\nu}\cal{P}^\alpha  \nonumber \\
&=& V(\tilde{t})T_{\mu\alpha\nu}\epsilon_\gamma^\mu\epsilon_\psi^{\nu}\cal{P}^\alpha\label{h6}
\end{eqnarray}
with
\begin{eqnarray}
T^{\mu\alpha\nu} &=& (k_1+k_2)^\alpha g^{\mu\nu} -2k_1^\nu g^{\alpha\mu}+2\left[k_1^\mu g^{\alpha\nu}\right. \nonumber\\
&& \left. +\frac{k_2^\nu}{k_2^2}(k_1\cdot k_2 g^{\alpha\mu}-k_1^\alpha k_2^\mu-k_1^\mu k_2^\alpha)\right. \nonumber\\
&& \left. -\frac{k_1^2 k_2^\mu}{k_2^2 k_1\cdot k_2}(k_2^2 g^{\alpha\nu}-k_2^\alpha k_2^\nu)\right]+(k_1-k_2)^\alpha g^{\mu\nu}.
\end{eqnarray}
In Eq. (\ref{h6}), $\beta_c$ is the effective coupling constant between a Pomeron and
a charm quark within $J/\psi$, and $\mu_0$ is a cutoff in the form
factor related to the Pomeron.

Thus, we get the amplitudes of Fig.~\ref{fig:2to3bg}~(a) and (b), i.e.,
\begin{eqnarray}
\mathcal{T}^{{Pomeron}}_{fi\,\,(a)} &=& g_{NN\omega} F_{pp\omega}(q_{s}^2)f(\tilde{t})V(\tilde{t})\mathcal{G}_P(s,\tilde{t})T^{\mu\rho\nu}
\epsilon_{\gamma}^{\mu}\epsilon_{\psi}^{\ast\nu}\epsilon_{\omega}^{\ast\alpha}\nonumber
\end{eqnarray}
\begin{eqnarray}
&& \times\bar{u}(p_2) \left(\gamma_\alpha + \frac{i\kappa_\omega}{2m_N}\sigma_{\alpha\beta}k_3^\beta\right) \frac{\slashed{q}_{s} + m_N}{q_{s}^2 - m_N^2}
\gamma^{\rho} u(p_1), \nonumber\\\label{h1}\\
\mathcal{T}^{{Pomeron}}_{fi\,\,(b)} &=& g_{NN\omega} F_{pp\omega}(q_{u}^2)f(\tilde{t})V(\tilde{t})\mathcal{G}_P(s,\tilde{t})T^{\mu\alpha\nu}
\epsilon_{\gamma}^{\mu}\epsilon_{\psi}^{\ast\nu}\epsilon_{\omega}^{\ast\alpha}\nonumber\\
&& \times\bar{u}(p_2) \gamma^{\rho} \frac{\slashed{q}_{u} + m_N}{q_{u}^2 - m_N^2}
\left(\gamma_\alpha + \frac{i\kappa_\omega}{2m_N}\sigma_{\alpha\beta}k_3^\beta\right) u(p_1)\nonumber\\\label{h2}
\end{eqnarray}
with $\mathcal{G}_P(s,\tilde{t})= -i(\alpha^\prime s)^{\alpha(\tilde{t})-1} $, which
is related to the Pomeron trajectory
$\alpha(\tilde{t})=1+\epsilon+\alpha^\prime \tilde{t}$.
The values of the parameters involved in the above amplitudes include: $\beta_0^2=4.0 \ \mbox{GeV}^2$, $\beta_c^2=0.8\ \mbox{GeV}^2$, $\alpha^\prime=0.25\ \mbox{GeV}^{-2}$, $\epsilon=0.08$, and $\mu_0=1.2\ \mbox{GeV}$.

To describe the effect of the off-shell intermediate nucleon as shown in Fig. \ref{fig:2to3bg}, a form factor for
the $pp\omega$ vertex is considered with the form $F_{pp\omega}(q^2)=(\Lambda_{pp\omega}^2-m_N^2)/(\Lambda_{pp\omega}^2 - q^2)$,
where $q^2$ is the momentum of the intermediate nucleon. Later, we will discuss how to constrain the value of the cutoff $\Lambda_{pp\omega}$.

With the amplitudes listed in Eq. (\ref{h0}) and Eqs. (\ref{h1})-(\ref{h2}), we obtain
the square of the total invariant transition amplitude
\begin{eqnarray}
|{\cal M}|^2 &=& \sum \left|T_{fi}^{Signal}+T_{fi\,\,(a)}^{Pomeron} +
T_{fi(\,\,b)}^{Pomeron}\right|^2.\label{2}
\end{eqnarray}
The corresponding total cross
section of the process $\gamma p\to J/\psi\omega p$ is
\begin{eqnarray}
 d\sigma &=& \frac{m_N^2}{|k_1\cdot p_1|}\frac{|\mathcal{M}|^2}{4}(2\pi)^4d\Phi_3(k_1+p_1;p_2,k_2,k_3),\label{1}
\end{eqnarray}
where the $n$-body phase space is defined as \cite{Beringer:1900zz}
$$d\Phi_n(P;p_1,...,p_n)=\delta^4(P-\sum\limits_{i=1}^np_i)\prod\limits_{i=1}^3\frac{d^3p_i}{(2\pi)^32E_i}.$$

\begin{figure}[htb]
\begin{center}
\scalebox{0.9}{\includegraphics[width=\columnwidth]{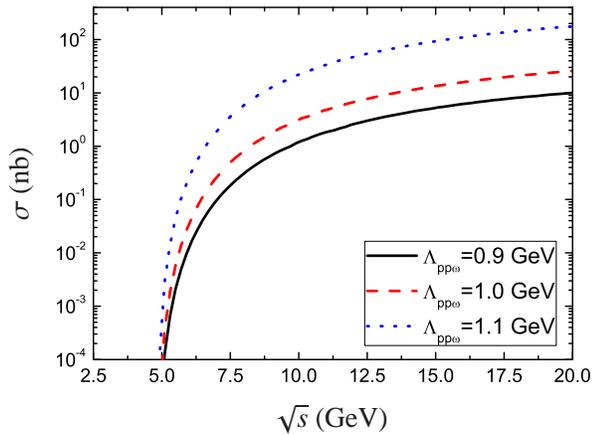}}
\caption{(color online). The total cross section of $\gamma p \to J/\psi\omega p$  from the Pomeron exchange. Here, we take different typical cutoff $\Lambda_{pp\omega}$ as input.
\label{fig:Pom-cutoff}}
\end{center}
\end{figure}

For numerically calculating the  total cross section including both signal and background contributions, we use the FOWL code. In Fig. \ref{fig:Pom-cutoff} the total
cross section of the Pomeron exchange contributions for $\gamma p\to J/\psi\omega p$ is given with
different typical $\Lambda_{pp\omega}$ values $(\Lambda_{pp\omega}=$ 0.9, 1.0 and 1.1 GeV).  Our calculation shows that the results are sensitive to the values of the cutoff. And the line shape of the total cross section dependent on $\sqrt{s}$ is monotonously increasing.
We notice that there were several experiments of $\gamma p\to J/\psi p$, where the measured cross section of this process is about 10 $nb$ \cite{JLAB,Levy:2007fb}. To some extend, the discussed $\gamma p\to J/\psi \omega p$ process is similar to $\gamma p\to J/\psi p$. It is natural to assume that the cross section of $\gamma p\to J/\psi \omega p$ should be similar or even lower than that of $\gamma p\to J/\psi p$. Under this assumption, we can roughly constrain the cutoff to be $\Lambda_{pp\omega}=0.9$ GeV, which will be applied to the following discussions.

\begin{figure}[htb]
\begin{center}
\scalebox{0.9}{\includegraphics[width=\columnwidth]{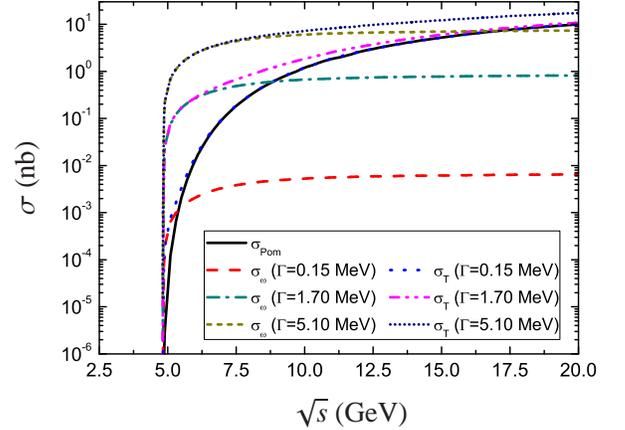}}
\caption{(color online). The energy dependence of the total cross sections for $\gamma p \to J/\psi\omega p$. Here, $\sigma_{Pom}$ and $\sigma_\omega$ are the results only considering the Pomeron exchange and the $\omega$ exchange contributions, respectively, while $\sigma_T$ denotes the total cross section of $\gamma p \to J/\psi\omega p$. We also give the variations of $\sigma_\omega$ and $\sigma_T$ to $\sqrt{s}$ corresponding to the typical values $\Gamma(X(3915)\to J/\psi\omega)=0.15$ MeV, $\Gamma(X(3915)\to J/\psi\omega)=1.7$ MeV and $\Gamma(X(3915)\to J/\psi\omega)=5.1$ MeV.
\label{fig:2to3res}}
\end{center}
\end{figure}

In Fig. \ref{fig:2to3res}, we further present the total cross section of $\gamma p \to J/\psi\omega p$ including the signal and background contributions, which corresponds to Eqs. (\ref{2})-(\ref{1}). By the comparison of the typical $\sigma_\omega$, $\sigma_T$  and $\sigma_{Pom}$, we find:

1. When $\Gamma(X(3915)\to J/\psi\omega)=0.15$ MeV, the corresponding $\sigma_\omega$ is rather small. However, $\sigma_\omega$ is larger than $\sigma_{Pom}$ around $\sqrt{s}=5$ GeV, which is close to the threshold of $\sqrt{s}$ of the $\gamma p \to J/\psi\omega p$ reaction.
Since the decay width of $X(3915)\to J/\psi\omega$ determines whether the $X(3915)$ signal is buried by the background,  the above analysis makes us obtain the lower limit of $\Gamma(X(3915)\to J/\psi\omega)$ if distinguishing the $X(3915)$ signal from the background.

2.  When $\Gamma(X(3915)\to J/\psi\omega)=1.7$ MeV and 5.1 MeV, the corresponding $\sigma_T$ is dominated by the $X(3915)$ signal contribution at $\sqrt{s}\leq 8$ and $\sqrt{s}\leq 15$ GeV, respectively (see Fig. \ref{fig:2to3res} for the detail). It means that the $X(3915)$ signal can be easily observed if taking suitable $\sqrt{s}$ range.

\begin{figure*}[htbp]
\begin{center}
\scalebox{1.0}{\includegraphics[width=\textwidth]{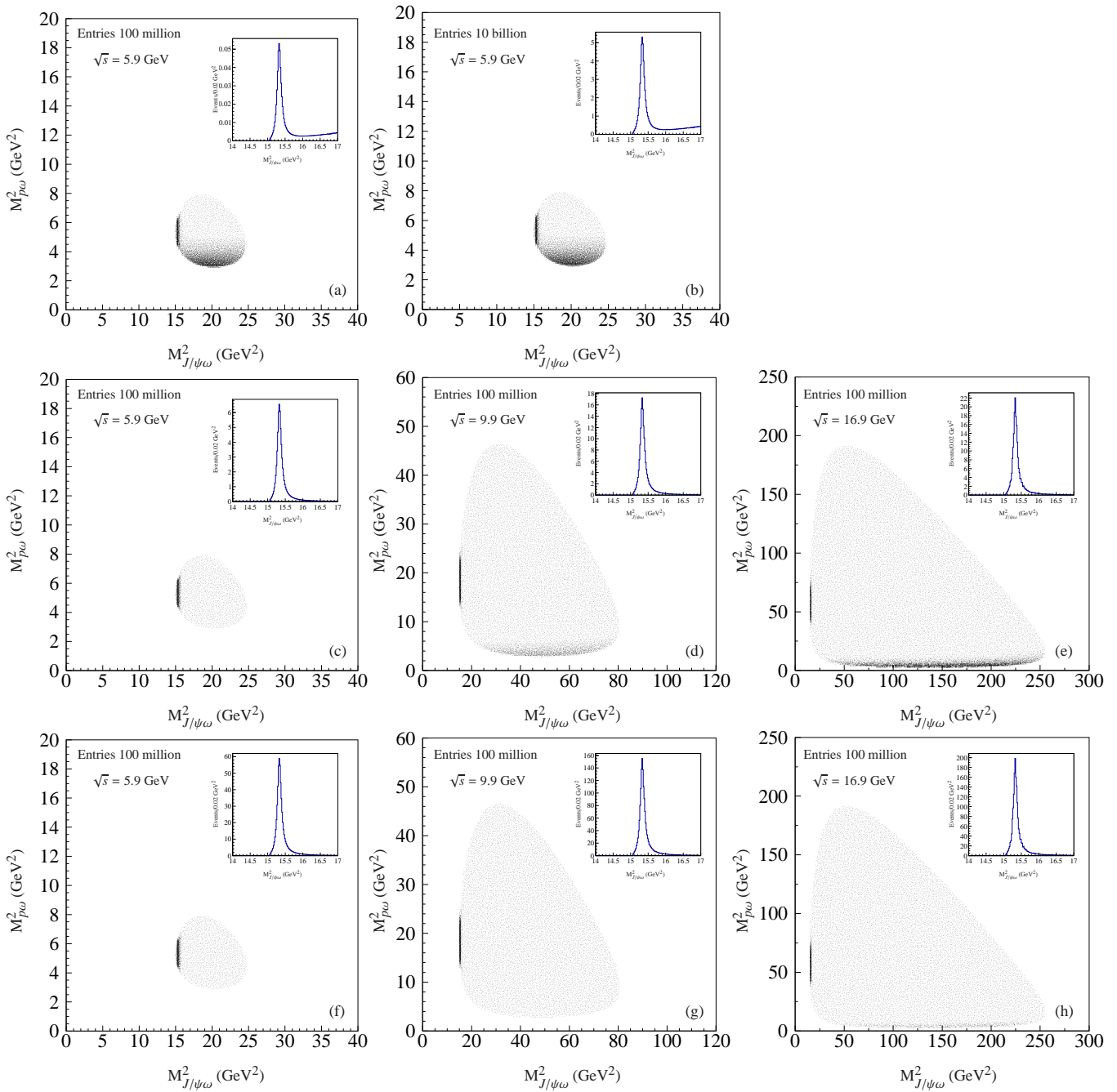}}
\caption{The Dalitz plot and the corresponding $J/\psi\omega$ invariant mass spectrum
for the $\gamma p \to J/\psi\omega p$ process. Here, the results listed in the first, the second and the third rows correspond to the typical value
$\Gamma(X(3915) \to J/\psi\omega)=0.15$ MeV, $\Gamma(X(3915) \to J/\psi\omega)=1.7$ MeV and $\Gamma(X(3915) \to J/\psi\omega)=5.1$ MeV, respectively.
\label{fig:DalitzPlot}}
\end{center}
\end{figure*}

Besides presenting the above study of the total cross section of $\gamma p \to J/\psi\omega p$,
we further carry out the analysis of the Dalitz plot, which can provide important information of the discovery potential of $X(3915)$ by photoproduction process $\gamma p \to J/\psi\omega p$.
In Fig. \ref{fig:DalitzPlot}, the Dalitz plot and the corresponding $J/\psi\omega$ mass spectrum with several typical values of $\sqrt{s}$ are listed. The numerical results indicate:

1. For the case with typical value $\Gamma(X(3915) \to J/\psi\omega)=0.15$ MeV,
it is difficult to identify $X(3915)$ signal if only taking 100 million collisions, since there is no enough signal event.
More collisions is helpful to the search for the $X(3915)$ signal (see the first row in Fig. \ref{fig:DalitzPlot}). {When performing a Monte Carlo simulation with 10 billion collisions, the signal event is 5 events/0.02 GeV$^{-2}$.} We need to emphasize that the ideal $\sqrt{s}$ range is around the threshold of the $\gamma p\to J/\psi\omega p$ reaction.

2. If $\Gamma(X(3915) \to J/\psi\omega)=1.7$ MeV, the signal of $X(3915)$ with $\sqrt{s}=9.9$ GeV is more explicit than that with $\sqrt{s}=5.9$ GeV. Then, with increasing $\sqrt{s}$, the background contribution becomes visible, which is the reason why there exists a horizontal band in Fig. \ref{fig:DalitzPlot} (d) and (e). This phenomena is consistent with the result from Fig. \ref{fig:2to3res} since the cross section from the Pomeron exchange goes up continuously. The situations corresponding to $\Gamma(X(3915) \to J/\psi\omega)=1.7$ MeV and $\Gamma(X(3915) \to J/\psi\omega)=5.1$ MeV are similar to each other. By the $J/\psi\omega$ invariant mass spectrum, we estimate that the event of $X(3915)$ reaches up to around 200/0.02 GeV$^{-2}$ at $\sqrt{s}=16.9$ GeV and with 100 million collisions, when $\Gamma(X(3915) \to J/\psi\omega)=5.1$ MeV is adopted.

3. The comparison of the results listed in the second and the third rows of Fig. \ref{fig:DalitzPlot} indicates that larger value of $\Gamma(X(3915) \to J/\psi\omega)$ is more beneficial to the search for $X(3915)$.

\begin{figure}[htb]
\begin{center}
\scalebox{0.9}{\includegraphics[width=\columnwidth]{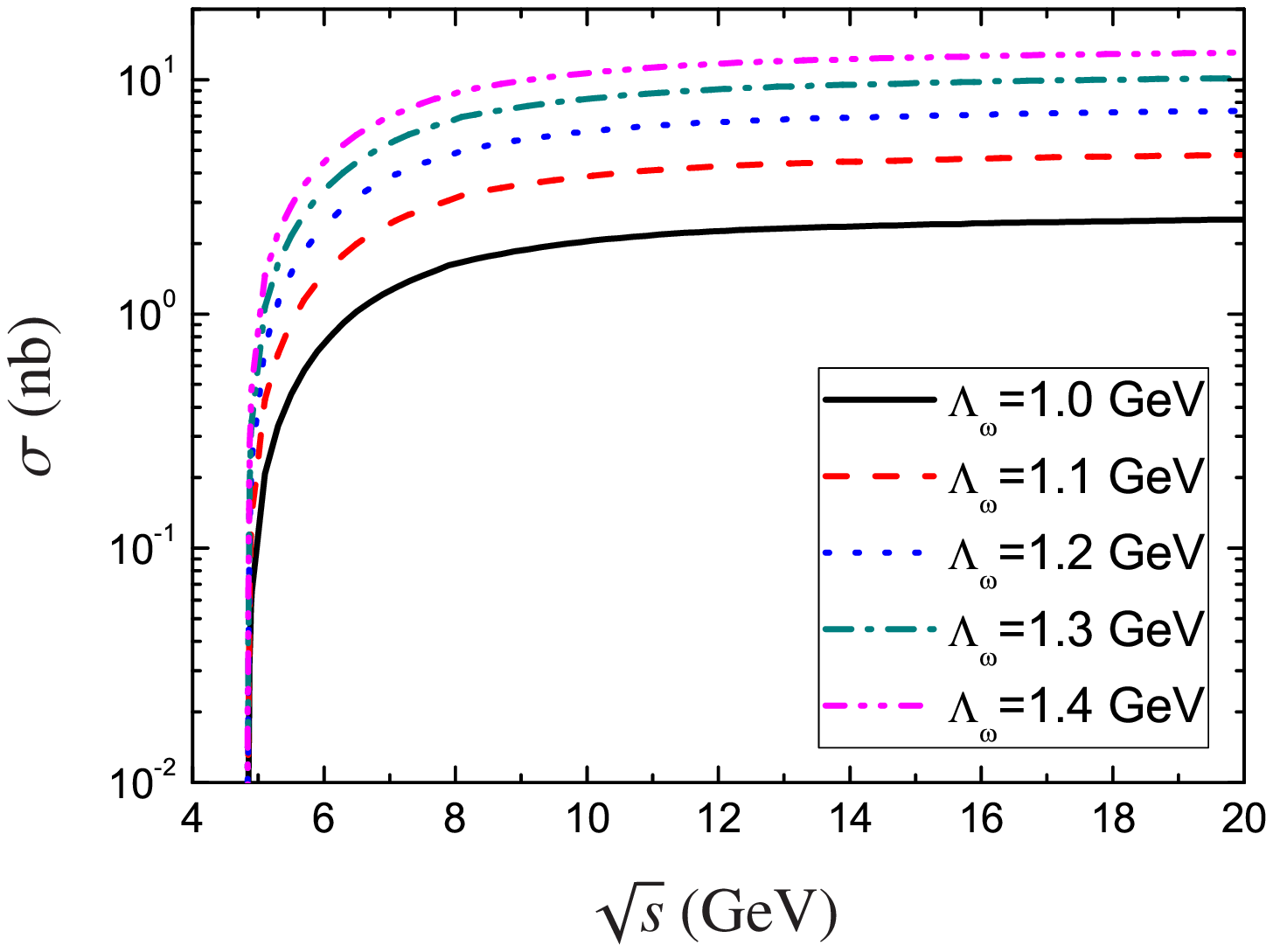}}
\caption{(color online). The total cross section of $\gamma p \to J/\psi\omega p$ from the signal contribution dependent on $\Lambda_\omega$. Here, the partial decay width is $\Gamma(X(3915)\to J/\psi\omega)=5.1$ MeV and the coupling constants are $(g_{NN\omega},\kappa_\omega)=(15.9,0)$.
\label{fig:Sig-cutoff}}
\end{center}
\end{figure}

In the above discussion, the cutoff $\Lambda_\omega$ for the $pp\omega$ vertex with off-shell $\omega$ meson is set to be 1.2 GeV as adopted in \cite{Gasparyan:2003fp}. In the following, we further give the dependence of the signal strength on the cutoff. In Fig. \ref{fig:Sig-cutoff}, we present the the $\Lambda_\omega$ dependence of the cross section of $\gamma p \to J/\psi\omega p$ from the signal contribution, where we take $\Lambda_\omega=1.0-1.4$ GeV with step of 0.1 GeV. As shown in Fig. \ref{fig:Sig-cutoff}, the line shape of the cross section are quite similar to that listed in Fig. \ref{fig:DSDT2to2}. The cross section is $2.5 \sim 13$ $nb$ corresponding to $\Lambda_\omega=1.0-1.4$ GeV.

\section{summary}\label{sec5}

In this work, we explore the discovery potential of charmonium-like state $X(3915)$ by meson photoproduction. As a good candidate of $\chi_{c0}^{\prime}(2P)$ \cite{Liu:2009fe}, $X(3915)$ was only observed in the $\gamma\gamma$ fusion process \cite{:2009tx}. Thus, searching for other processes to investigate $X(3915)$ is an interesting and important topic, which will be valuable to further deepen our understanding of $X(3915)$.

Since the final state of the observed $X(3915)$ decay contain two vector mesons, meson photoproduction process
can be suitable to study $X(3915)$. For quantitatively answering whether $X(3915)$ can be observed
in meson photoproduction process, we study the $\gamma p \to J/\psi\omega p$ process by including the $X(3915)$
signal and background contributions. Furthermore, the corresponding cross section
and the analysis of the Dalitz plot are given, which provide abundant information to the experimental study of
$X(3915)$ by meson photoproduction.

Our study also shows that the experimental measurement of the decay width of $X(3915)\to J/\psi\omega$ is a crucial input in studying the meson photoprodution of $X(3915)$. However, this key value is still absent in experiments. In this work, we consider several typical values of $\Gamma(X(3915)\to J/\psi\omega)$ to discuss the discovery potential of $X(3915)$ by meson photoproduction. We expect further experiment of the hidden-charm decay of $X(3915)$. The following theoretical and experimental joint effort will be helpful to probing charmonium-like state $X(3915)$ through meson photoproduction.

\vfil
\section{Acknowledgements}
This project is supported
by the National Natural Science Foundation of China under
Grants No. 11222547, No. 11175073, and No. 11035006,
the Ministry of Education of China (SRFDP under Grant No. 20120211110002 and the Fundamental Research Funds for the Central
Universities), and the Fok Ying-Tong Education
Foundation (No. 131006).

\vfill

\end{document}